\documentstyle[12pt,rotate]{article}

\input epsf.sty

\newcommand{\inclfig}[2]{\mbox{\epsfxsize=#1cm \epsfbox{#2.ps}}}

\newcommand{\g}{{\sl g}}

\def\1{\hbox{{1}\kern-.25em\hbox{l}}}

\begin{document}

\begin{titlepage}

\centerline{\large \bf Evolution of non-forward parton distributions }
\centerline{\large \bf in next-to-leading order: singlet sector.}

\vspace{15mm}

\centerline{\bf A.V. Belitsky\footnote{Alexander von Humboldt Fellow.},
            D. M\"uller, L. Niedermeier, A. Sch\"afer}

\vspace{15mm}

\centerline{\it Institut f\"ur Theoretische Physik, Universit\"at
                Regensburg}
\centerline{\it D-93040 Regensburg, Germany}

\vspace{20mm}

\centerline{\bf Abstract}

\hspace{0.5cm}

We present a general method for the solution of the renormalization
group equations for the non-forward parton distributions on the
two-loop level in the flavour singlet channel based on an orthogonal
polynomial reconstruction. Using this formalism we study the effects
of the evolution on recently proposed model distribution functions.

\vspace{7cm}


\vspace{0.5cm}


\end{titlepage}

%%%%%%%%%%%%%%%%%%%%%%%%%%%%%%%%%%%%%%%%%%%%%%%%%%%%%%%%%%%%%%%%%%%%%
\section{Introduction.}
%%%%%%%%%%%%%%%%%%%%%%%%%%%%%%%%%%%%%%%%%%%%%%%%%%%%%%%%%%%%%%%%%%%%%

Parton distribution functions as gained from high-energy scattering
reactions provide our main information about the non-perturbative
structure of hadrons. Up to now the wealth of this knowledge was
obtained from deep inelastic scattering (DIS) of leptons off a hadron
target which allows to access the ordinary forward structure functions.
Recently much excitement was generated by new objects which could provide
a new insight into the underlying dynamics of non-perturbative QCD
in hard processes --- the non-forward parton distributions
\cite{Ji98,Rad97,DitMulGeyRobHor94}. Several processes were proposed
where the latter could be measured, e.g. deeply virtual Compton
scattering, diffractive production of vector meson etc. It is
unnecessary to emphasize that any precise analysis of such data will
require accurate predictions from strong interaction theory for the
corresponding reactions, especially for the radiative corrections to
the tree level amplitudes. We solved the latter problem in a series
of our recent papers (see \cite{BelMueNidSch98} and references cited
therein) where all necessary ingredients for a next-to-leading order
(NLO) analysis were derived. In this contribution
we conclude by presenting a general method for the solution of the
generalized evolution equations for the non-forward parton distributions
in two-loop approximation in the flavour singlet channel. The technique
we use, i.e. the reconstruction of a distribution function\footnote{The
functions ${\cal O} (x, \zeta)$ used here vanishes outside the region
$x \in [0, 1]$. However, below we deal with non-forward parton
distributions with the support $-1 + \zeta \leq x \leq 1$ which
parametrize the matrix element of a light-ray operator. The results for
the former case can be found in Appendix \ref{CoeffNFPD}.}, ${\cal O}
(x, \zeta)$, from its conformal moments, ${\cal O}_j$, via an expansion
in the series of the orthogonal Jacobi polynomials \cite{BatErd53},
$P^{(\alpha,\beta)}_j$,
\begin{equation}
\label{SampleExp}
{\cal O} (x, \zeta)
\propto w (x|\alpha, \beta) \sum_{j}^{N_{\rm max}}
P^{(\alpha,\beta)}_j (2x - 1) \sum_{k}^{j} E_{jk} (\zeta)\, {\cal O}_k ,
\end{equation}
is close in spirit to the one developed about two decades ago by
Parisi and Sourlas \cite{ParSou79} (see also Ref.\ \cite{BarLanSna81})
for the usual DIS. The generalization to the singlet sector in DIS was
done in Ref.\ \cite{Kri87}. Quite recently it was independently
rediscovered by us in Refs.\ \cite{BelMueNidSch98,BelGeyMueSch97}
in relation to the problem at hand. This approach is the most
suitable for our purposes because the standard methods of
contour integration via the inverse Mellin transformation
are very hard to implement in practice. On top of this for
particular values for the parameters of the Jacobi polynomials the latter
become eigenfunctions of the leading order non-forward evolution
equation which favours the choice we have made. Since for the time being
analytical expressions in NLO exist only for the complete set of
non-forward singlet anomalous dimensions \cite{BelMue98a,BelMue98b} but
not for the kernels themselves it is the only possible method to go beyond
the one-loop approximation. Though, using the generating function for the
diagonal part of the non-forward evolution kernels derived and tested at
leading order in Ref.\ \cite{BelMue98a} (the non-diagonal parts are known
analytically) one can extend this procedure to two-loop approximation
and find in principle NLO exclusive kernels, $V (t, t^\prime)$ with
$|t|,|t^\prime| \leq 1$, analytically. After continuation to the entire
plane $\{ |t|,|t^\prime| < \infty \}$ might allow a direct numerical
integration of the two-loop evolution equations --- an alternative
method used in the analysis of the forward structure functions.

Note, however, that with the former approach at hand, our numerical
task is more difficult contrary to method used in usual forward
scattering as we cannot restrict ourselves to the first
$N_{\rm max} \sim 10$ polynomials in the series (\ref{SampleExp})
because in DIS the shape of the curve is roughly fixed by the
weight-function, $w(x|\alpha, \beta) = \bar x^\alpha x^\beta$
(with $\bar x \equiv 1 - x$), provided we have made an appropriate choice
for the indices of the Jacobi polynomials, namely, with $\beta$ given
by Regge theory predictions and $\alpha$ driven by quark counting rules
near the phase space boundary\footnote{Let us remind, however, that there
exists another optimal choice of $\alpha$ and $\beta$ which leads
to the same high precision reconstruction of the structure function
\cite{Kri87}. Obviously, there is no dependence on the particular values
of these parameters provided a large enough number of terms, $N_{\rm max}$,
is taken into account.}. The series of polynomials leads only to small
perturbations around this $\bar x^\alpha x^\beta$-behaviour. For the case
at hand the situation is different since the shape of the distribution
is obtained from the total series over all oscillating polynomials.
In practice, due to rather rapid convergence of the series the number
of terms to be taken into account is rather large but still treatable.

In the present paper we pursue the goal of developing a machinery for
the solution of the two-loop renormalization group equations for the
non-forward parton distributions and of studying numerically the
$Q^2$-dependence of the latter. Our presentation is organized as follows.
In the next section we give our definitions and conventions used
throughout the paper. The third section is devoted to
the description of our formalism for evolution of generalized distribution
functions together with a set of explicit formulae required in the
numerical analysis. Since there are no experimental data yet for the
non-forward parton distributions we have to rely on estimations of the
form of the $(x,\zeta)$-dependence at low normalization scale from
non-perturbative approaches to QCD. In section 4 we give our results
for models of the singlet non-forward parton functions. We consider
CTEQ-based functions which are deduced making plausible
assumptions about the behaviour of the so-called double distributions
\cite{Rad97} in different regions of phase space. The final section
is left for the conclusions. To make the presentation self-consistent as
much as possible we add three appendices. The first one contains the
analytically continued forward anomalous dimension matrix responsible
for the evolution of the multiplicatively renormalizable conformal
moments. The second one presents the non-diagonal elements of the
anomalous dimension matrix of the tree-level conformal operators which
define the corrections to the eigenfunctions of the NLO evolution
kernels. The last appendix contains the formulae required for
reconstruction of the non-forward distributions with the support
$0 \leq x \leq 1$.

%%%%%%%%%%%%%%%%%%%%%%%%%%%%%%%%%%%%%%%%%%%%%%%%%%%%%%%%%%%%%%%%%%%%%
\section{Conventions.}
%%%%%%%%%%%%%%%%%%%%%%%%%%%%%%%%%%%%%%%%%%%%%%%%%%%%%%%%%%%%%%%%%%%%%

Since we are interested in the study of the parity even flavour singlet
evolution equations we face as usual the mixing problem between quarks
and gluons. To treat it in a compact way let us introduce a two-dimensional
vector of singlet non-forward parton distributions composed from quark and
gluon functions ($\bar\zeta \equiv 1 - \zeta$):
\begin{equation}
\mbox{\boldmath$\cal O$} (x, \zeta)
= \left( \, { {\cal Q} (x, \zeta) \atop {\cal G} (x, \zeta)} \, \right)
\qquad\mbox{with}\qquad
- \bar\zeta \leq x \leq 1.
\end{equation}
The latter are defined as the light-cone Fourier transforms\footnote{Below,
we will repeatedly omit the integration limits. The latter can be easily
restored making use of known support properties of the distributions
or/and evolution kernels.}
\begin{eqnarray}
\label{NFPD-Q}
\langle h^\prime |
{^Q\!{\cal O}} (\kappa_1, \kappa_2)
| h \rangle
&=& 2 \int_{- \bar\zeta}^{1} dx \
e^{- i \kappa_1 x - i \kappa_2 (\zeta - x)}
{\cal Q} (x, \zeta),\\
\label{NFPD-G}
\langle h^\prime |
{^G\!{\cal O}} (\kappa_1, \kappa_2)
| h \rangle
&=& \int_{- \bar\zeta}^{1} dx \
e^{- i \kappa_1 x - i \kappa_2 (\zeta - x)}
{\cal G} (x, \zeta),
\end{eqnarray}
of the light-ray quark and gluon string operators ($v_+ \equiv
v_\mu n_\mu$)
\begin{equation}
\label{treeLRO}
{^Q\!{\cal O}} (\kappa_1,\kappa_2)
=
\bar{\psi}(\kappa_2 n)
\gamma_+
\psi (\kappa_1 n)
-
\bar{\psi}(\kappa_1 n)
\gamma_+
\psi(\kappa_2 n) , \quad
{^G\!{\cal O}}
(\kappa_1,\kappa_2)
=
G_{+ \mu} (\kappa_2 n)
G_{\mu +} (\kappa_1 n) ,
\end{equation}
where, for brevity, we omit a path-ordered link factor which ensures
gauge invariance. Here we accept the conventions advocated by Radyushkin
\cite{Rad97}, namely, $x$ is the momentum fraction of an outgoing
parton w.r.t.\ the incoming hadron momentum $p$, $k_+ = x p_+$, and
$\Delta_+ \equiv p_+ - p^\prime_+ = \zeta p_+$, where $p'$ is an outgoing
hadron momentum. These quantities are related to the variables $t$ and
$\eta$ used by the authors \cite{Ji98,DitMulGeyRobHor94} according to
$\eta = \frac{\zeta}{2 - \zeta}$, $t = \frac{2x - \zeta}{2 - \zeta}$,
where $k_+ = t \bar P_+$, $\Delta_+ = \eta \bar P_+$ and the averaged
momentum is introduced as follows $\bar P = p + p'$. An advantage of
the first conventions is that the variable $x$ acquires a simple partonic
interpretation in contrast to $t$. However, the range of the latter does
not depend on the longitudinal asymmetry parameter $\eta$ as compared
to $x$.

Due to charge conjugation properties of the non-local operators
(\ref{treeLRO}) it immediately follows that in the $(x, \zeta)$-plane
the singlet quark distribution is anti-symmetric, ${\cal Q}
(x, \zeta) = - {\cal Q} (\zeta - x, \zeta)$, while the gluon one is
symmetric, ${\cal G} (x, \zeta) = {\cal G} (\zeta - x, \zeta)$,
w.r.t.\ the line $x = \frac{\zeta}{2}$. Up to an obvious redefinition the
functions ${\cal Q}$ and ${\cal G}$ coincide with the ones introduced
in Refs.\ \cite{Ji98,DitMulGeyRobHor94}
\begin{equation}
{\cal Q} \left( x = \frac{t + \eta}{1 + \eta},
\zeta = \frac{2 \eta}{1 + \eta} \right)
\equiv (1 + \eta) q (t, \eta),
\end{equation}
and similar for gluons, where the variables $t$ and $\eta$ vary
within the limits $- 1 \leq t \leq 1$, $0 \leq \eta \leq 1$. Obviously,
these singlet quark and gluon distributions are, respectively, odd
and even functions of $t$.

The original functions, ${\cal Q}$ and ${\cal G}$, introduced above can
be decomposed into the following quark, ${^{Q}\!{\cal O}}$, anti-quark,
${^{\bar Q}\!{\cal O}}$, and gluon, ${^{G}\!{\cal O}}$, non-forward
distributions defined in the range $x \in [0, 1]$ via
\begin{eqnarray}
\label{Q-decomp}
{\cal Q} (x, \zeta)
&=& \frac{1}{2} \left\{
\left[
{^{Q}\!{\cal O}} (x, \zeta)
+ {^{\bar Q}\!{\cal O}} (x, \zeta) \right.
\right] \theta (x) \theta (\bar x) \nonumber\\
&&\qquad\qquad\qquad\qquad - \left.\left[
{^{Q}\!{\cal O}} (\zeta - x, \zeta)
+ {^{\bar Q}\!{\cal O}} (\zeta - x, \zeta)
\right] \theta (\zeta - x) \theta (x + \bar\zeta)
\right\}, \\
\label{G-decomp}
{\cal G} (x, \zeta)
&=& \frac{1}{2} \left\{
{^{G}\!{\cal O}} (x, \zeta)
\theta (x) \theta (\bar x)
+
{^{G}\!{\cal O}} (\zeta - x, \zeta)
\theta (\zeta - x) \theta (x + \bar\zeta)
\right\} .
\end{eqnarray}
However, our consequent discussion deals entirely with the non-forward
distributions ${\cal Q}$ and ${\cal G}$.

%%%%%%%%%%%%%%%%%%%%%%%%%%%%%%%%%%%%%%%%%%%%%%%%%%%%%%%%%%%%%%%%%%%%%
\section{The method.}
%%%%%%%%%%%%%%%%%%%%%%%%%%%%%%%%%%%%%%%%%%%%%%%%%%%%%%%%%%%%%%%%%%%%%

Due to ultraviolet divergences of perturbative corrections for a
product of operators separated by a light-like distance as in Eq.\
(\ref{treeLRO}) the distributions acquire a dependence
on a normalization point, $\mu^2$, governed by a renormalization group
equation, the so-called generalized Efremov-Radyushkin-Brodsky-Lepage
(ER-BL) evolution equation
\begin{equation}
\label{EvEq}
\mu^2\frac{d}{d\mu^2} \mbox{\boldmath$\cal O$} (x, \zeta)
= \int d x^\prime \mbox{\boldmath$K$}
\left( x , x^\prime, \zeta \left| \alpha_s \right)\right.
\mbox{\boldmath$\cal O$} (x^\prime, \zeta) ,
\end{equation}
where the purely perturbative kernel $\mbox{\boldmath$K$}$ is a $2 \times 2$
matrix given by a series in the coupling constant, $\alpha_s$. The LO
non-forward light-cone fraction kernels\footnote{For recent independent
calculations of these kernels related to the problem at hand see the
reviews \cite{Ji98,Rad97}.} were evaluated many years ago in Ref.\
\cite{BFKL85} for even and odd parity and chirality while corresponding
light-cone position counterparts were addressed in Refs.\
\cite{Rad97,BraGeyRob87}-\cite{BelMue97a}. The evaluation of the
two-loop corrections, however, for the Gegenbauer moments of the
kernel $\mbox{\boldmath$K$}$
\begin{equation}
\label{MomGegen}
\int\! dx\, C^{\nu (A)}_j \left( 2\frac{x}{\zeta} - 1 \right)
{^{AB}\!K}
\left( x , x^\prime, \zeta \left| \alpha_s \right)\right.
= - \frac{1}{2}
\sum_{k=0}^j {^{AB}\!\gamma_{jk}}(\alpha_s) C^{\nu (B)}_k
\left(2 \frac{x^\prime}{\zeta} - 1 \right),
\end{equation}
--- where the numerical values of the indices $\nu (A,B)$ depends on
the channel under consideration --- has been addressed by us in
Refs.\ \cite{BelMue97a,BelMue98a,BelMue98b}. In one-loop approximation
the above kernel is diagonal $\gamma_{jk}^{(0)} =
\gamma_j^{{\rm D}(0)} \delta_{jk}$ in this basis while beyond LO
non-diagonal elements, $\gamma_{jk}^{\rm ND}$, appear $\gamma_{jk}
= \gamma_{jk}^{\rm D} + \gamma_{jk}^{\rm ND}$. The diagonal
anomalous dimensions coincide up to pre-factors given in Eq.\
(\ref{relations}) with the anomalous dimensions of local operators
without total derivatives which appear in the operator product expansion
(OPE) for DIS. The formalism we have developed there allowed us to find
all entries, $\gamma_{jk}^{\rm ND}$, in closed analytical form in NLO and
provided a simple diagonalization of the evolution equation (\ref{EvEq}).
The following discussion concerns the solution of Eq.\ (\ref{EvEq})
within the formalism we have sketched above.

%%%%%%%%%%%%%%%%%%%%%%%%%%%%%%%%%%%%%%%%%%%%%%%%%%%%%%%%%%%%%%%%%%%%%
%            Figure 1
%%%%%%%%%%%%%%%%%%%%%%%%%%%%%%%%%%%%%%%%%%%%%%%%%%%%%%%%%%%%%%%%%%%%%
\begin{figure}[t]
\begin{center}
\vspace{4.7cm}
\hspace{-2cm}
\mbox{
\begin{picture}(0,220)(270,0)
\put(0,-30)                    {
\epsffile{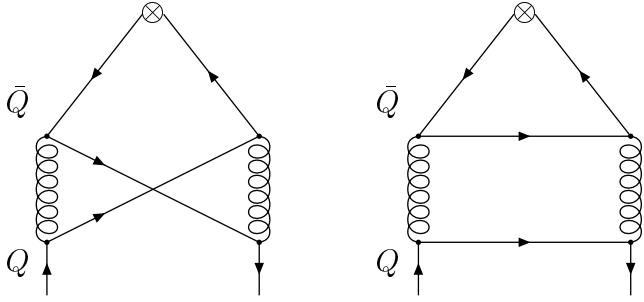}
                               }
\end{picture}
}
\end{center}
\vspace{-8.0cm}
\caption{\label{mixing} Two-loop diagrams giving rise to a non-vanishing
contribution to the ${\cal W}$-functions of the generalized ER-BL
evolution kernel (\protect\ref{ER-BL-Structure}).}
\end{figure}

Let us add a remark about the general structure of the all-order
evolution kernel $\frac{1}{|\eta|} \mbox{\boldmath$V$}
\left(\frac{t}{\eta}, \frac{t^\prime}{\eta} \right) = \mbox{\boldmath$K$}
\left( x = \frac{t + \eta}{1 + \eta}, x^\prime = \frac{t^\prime + \eta}{
1 + \eta}, \zeta = \frac{2 \eta}{1 + \eta} \right)$. It can be
deduced in a straightforward manner from the support properties and
charge conjugation symmetry of the light-ray operators and reads, e.g.
for the $QQ$-channel \cite{DitMueRobGeyHor88},
\begin{equation}
\label{ER-BL-Structure}
V(t, t^\prime) = \Theta_{11}^0 (t - t^\prime, t - 1)
{\cal V} (t, t^\prime)
+ \Theta_{11}^0 (- t - t^\prime, - t - 1)
{\cal W} (- t, t^\prime)
+ \left\{ { t \to - t \atop t^\prime \to - t^\prime } \right\} ,
\end{equation}
where we have introduced the step-function $\Theta_{11}^0 (t, t^\prime)
= \frac{1}{t - t^\prime}\left\{ \theta (t) \theta (- t^\prime)
- \theta (- t) \theta (t^\prime) \right\}$ \cite{BelMue97b} and
${\cal V},\ {\cal W}$ are analytic functions of their arguments.
Here the ${\cal W}$-part appears first in NLO and is generated
by diagrams depicted in Fig.\ \ref{mixing}. These terms give rise
to mixing of partons from different regions of phase space
conserved by the leading order evolution and it forces us to consider
the functions (\ref{Q-decomp},\ref{G-decomp}) which are defined on the
entire interval of $x \in [- \bar\zeta, 1]$.

%%%%%%%%%%%%%%%%%%%%%%%%%%%%%%%%%%%%%%%%%%%%%%%%%%%%%%%%%%%%%%%%%%%%%
\subsection{General formalism.}
%%%%%%%%%%%%%%%%%%%%%%%%%%%%%%%%%%%%%%%%%%%%%%%%%%%%%%%%%%%%%%%%%%%%%

In order to solve the evolution equation (\ref{EvEq}) one has to find
its eigenvalues and eigenfunctions. This problem has been exhaustively
treated by us in general form in Ref.\ \cite{BelMue98b} where it was
shown that its solution can be written in terms of the partial conformal
wave expansion
\begin{equation}
\label{ConfExp}
\mbox{\boldmath$\cal O$} (x, \zeta, Q^2)
= \sum_{j = 0}^{\infty}
\mbox{\boldmath$\phi$}_j
\left( x, \zeta \left| \alpha_s (Q^2) \right)\right.
\widetilde{\mbox{\boldmath$\cal O$}}_j ( \zeta, Q^2 ),
\end{equation}
with the partial conformal waves matrix
\begin{equation}
\label{ConfWaves}
\mbox{\boldmath$\phi$}_j
\left( x, \zeta \left| \alpha_s (Q^2) \right)\right.
= \sum_{k = j}^{\infty} \zeta^{k - j}
\mbox{\boldmath$\phi$}_k
\left( x, \zeta \right) \mbox{\boldmath$B$}_{kj},
\end{equation}
being the eigenstate of the all-orders equation (\ref{EvEq}). The
$\mbox{\boldmath$B$}$-matrix defines the corrections to the eigenfunctions
which at tree-level diagonalize the leading order ER-BL equations. The
latter read\footnote{Let us emphasize that here the Gegenbauer polynomials
should be understood as mathematical distributions (see Ref.\ \cite{Rad97})
according to the relation \cite{BelSch98} ${\scriptstyle \frac{1}{\zeta}}
w ({\scriptstyle \frac{x}{\zeta}} | \nu)
C^\nu_j \left( 2 {\scriptstyle \frac{x}{\zeta}} - 1 \right) \propto
\int_{0}^{1} dy (y \bar y)^{j + \nu - 1/2} \delta^{(j)} (\zeta y - x)$
in order to be able to restore the correct support properties of the
non-forward distributions.} ($w (x|\nu) = (x \bar x)^{\nu - 1/2}$)
\begin{equation}
\mbox{\boldmath$\phi$}_j \left( x, \zeta \right)
= \frac{1}{\zeta}
\left(
\begin{array}{cc}
\frac{1}{\zeta^j}
\frac{w ( \frac{x}{\zeta} | \frac{3}{2} )}{N_j (\frac{3}{2})}
C_j^{\frac{3}{2}} \left( 2 \frac{x}{\zeta} - 1 \right) & 0 \\
0 & \frac{1}{\zeta^{j - 1}}
\frac{w ( \frac{x}{\zeta} | \frac{5}{2} )}{N_{j - 1} (\frac{5}{2})}
C_{j - 1}^{\frac{5}{2}} \left( 2 \frac{x}{\zeta} - 1 \right)
\end{array}
\right) .
\end{equation}
The multiplicatively renormalizable moments,
$\widetilde{\mbox{\boldmath$\cal O$}}_j ( \zeta, Q^2 )$, evolve as
follows
\begin{equation}
\label{Solution}
\widetilde{\mbox{\boldmath$\cal O$}}_j ( \zeta, Q^2 )
= \mbox{\boldmath$\cal E$}_j
\left( \alpha_s (Q^2), \alpha_s (Q_0^2) \right)
\widetilde{\mbox{\boldmath$\cal O$}}_j ( \zeta, Q_0^2 ) ,
\end{equation}
with an evolution operator defined by the equation
\begin{equation}
\label{EvolOperator}
\mbox{\boldmath$\cal E$}_j \left( \alpha_s (Q^2), \alpha_s (Q_0^2) \right)
= {\cal T} \exp
\left\{
- \frac{1}{2} \int_{Q_0^2}^{Q^2} \frac{d \tau}{\tau}
\mbox{\boldmath$\gamma$}_j^{\rm D} \left( \alpha_s (\tau) \right)
\right\}
\end{equation}
where the operator ${\cal T}$ orders the matrices of the diagonal
anomalous dimensions
\begin{equation}
\mbox{\boldmath$\gamma$}_j^{\rm D}
=
\left(
\begin{array}{cc}
{^{QQ}\!{\gamma}^{\rm D}_j} & {^{QG}\!{\gamma}^{\rm D}_j} \\
{^{GQ}\!{\gamma}^{\rm D}_j} & {^{GG}\!{\gamma}^{\rm D}_j}
\end{array}
\right)
\end{equation}
along the integration path. They are given as an expansion in the
coupling constant by $\mbox{\boldmath$\gamma$}_j \left( \alpha_s \right)
= \frac{\alpha_s}{2 \pi} \ \mbox{\boldmath$\gamma$}_j^{(0)}
+ \left( \frac{\alpha_s}{2 \pi} \right)^2 \mbox{\boldmath$\gamma$}_j^{(1)}
+ \dots$ ---  The elements of this matrix
are related to the forward anomalous dimensions $\gamma^{\rm fw}_j$
via the relations
\begin{equation}
\label{relations}
{^{QQ}\!{\gamma}^{\rm D}_j} = {^{QQ}\!{\gamma}^{\rm fw}_j} ,\quad
{^{QG}\!{\gamma}^{\rm D}_j} = \frac{6}{j}\ {^{QG}\!{\gamma}^{\rm fw}_j} ,
\quad
{^{GQ}\!{\gamma}^{\rm D}_j} = \frac{j}{6}\ {^{GQ}\!{\gamma}^{\rm fw}_j} ,
\quad
{^{GG}\!{\gamma}^{\rm D}_j} = {^{GG}\!{\gamma}^{\rm fw}_j} .
\end{equation}
The pre-factors $6/j,\ j/6$ in the off-diagonal matrix elements come from
the conventional definition of the Gegenbauer polynomials.

The normalization condition in Eqs.\ (\ref{ConfExp},\ref{ConfWaves}) is
chosen so that there are no radiative corrections at an input scale $Q_0$.
Apart from the minimization of the higher order corrections the
advantage of this choice lies in the fact that multiplicatively
renormalizable $\widetilde{\mbox{\boldmath$\cal O$}}_j ( \zeta, Q_0^2 )$
are defined at $Q_0$ by forming ordinary Gegenbauer moments with
non-forward distributions via
\begin{equation}
\widetilde{\mbox{\boldmath$\cal O$}}_j ( \zeta, Q_0^2 )
= \int\! dx\, {\mbox{\boldmath$C$}}_j ( x, \zeta )
{\mbox{\boldmath$\cal O$}} ( x, \zeta, Q_0^2 ), \ \mbox{with} \
{\mbox{\boldmath$C$}}_j ( x, \zeta )
=
\left(
\begin{array}{cc}
\zeta^j C_j^{\frac{3}{2}} \left( 2\frac{x}{\zeta} - 1 \right) & 0 \\
0 &
\zeta^{j - 1} C_{j - 1}^{\frac{5}{2}} \left( 2 \frac{x}{\zeta} - 1 \right)
\end{array}
\right).
\end{equation}

As before \cite{BelMueNidSch98} the restoration of the support
properties of the distributions is achieved via an expansion of
the latter in a series with respect to the complete set of
orthogonal polynomials, ${\cal P}^{( \alpha_p )}_j (t)$, on the
interval $- 1 \leq t \leq 1$
\begin{equation}
\label{PolySeries}
{\mbox{\boldmath$\cal O$}} ( x, \zeta, Q^2 )
= \frac{2}{2 - \zeta}
\sum_{j = 0}^{\infty}
\widetilde{\mbox{\boldmath$\cal P$}}_j
\left( \frac{2 x - \zeta}{2 - \zeta} \right)
{\mbox{\boldmath$\cal M$}}_j ( \zeta, Q^2 ),
\end{equation}
\begin{equation}
\label{PolyMatrix}
\widetilde{\mbox{\boldmath$\cal P$}}_j ( t )
=
\left(
\begin{array}{cc}
\frac{w (t | \alpha_p )}{n_j ( \alpha_p )}
{\cal P}_j^{( \alpha_p )} (t)
& 0 \\
0 &
\frac{w (t | \alpha_p^\prime )}{
n_j ( \alpha_p^\prime )}
{\cal P}_j^{( \alpha_p^\prime )} (t)
\end{array}
\right)
\end{equation}
with $w (t | \alpha_p )$ and $n_j ( \alpha_p )$ being weight
and normalization factors, respectively (see, e.g. \cite{AbrSte65}).
The matrix of moments, ${\mbox{\boldmath$\cal M$}}_j ( \zeta, Q^2 )$,
is given by the sum
\begin{eqnarray}
\label{JacobiMom}
{\mbox{\boldmath$\cal M$}}_j ( \zeta, Q^2 )
&=& \sum_{k = 0}^{\infty}
{\mbox{\boldmath$E$}}_{jk} (\zeta)
{\mbox{\boldmath$\cal O$}}_k (\zeta, Q^2), \ \mbox{where} \\
{\mbox{\boldmath$\cal O$}}_j (\zeta, Q^2)
&=& \sum_{k = 0}^{j} \zeta^{j - k}
{\mbox{\boldmath$B$}}_{jk}
\widetilde{\mbox{\boldmath$\cal O$}}_k ( \zeta, Q^2 ),
\end{eqnarray}
and the upper limit in Eq.\ (\ref{JacobiMom}) will come from the
constraint $\theta$-functions in the expansion coefficients
\begin{equation}
\label{E-matrix}
{\mbox{\boldmath$E$}}_{jk} (\zeta)
=
\left(
\begin{array}{cc}
E_{jk} (\frac{3}{2}; \alpha_p | \zeta) & 0 \\
0 & E_{j\,k-1} (\frac{5}{2}; \alpha_p^\prime | \zeta)
\end{array}
\right),
\end{equation}
with elements given by the integral (where $\theta_{jk}
= \{1,\ \mbox{if}\ j \geq k;\ 0,\ \mbox{if}\ j < k \}$)
\begin{eqnarray}
E_{jk} (\nu; \alpha_p | \zeta) = \frac{\theta_{jk}}{(2\zeta)^k}
\frac{\Gamma (\nu) (\nu + k)}{
\Gamma (\frac{1}{2}) \Gamma (k + \nu + \frac{1}{2})}
\int_{- 1}^{1} dt\ (1 - t^2)^{k + \nu - \frac{1}{2}}
\frac{d^k}{dt^k}\ {\cal P}^{( \alpha_p )}_j
\left( \frac{\zeta t}{2 - \zeta} \right).
\end{eqnarray}
Taking Jacobi polynomials, ${\cal P}^{( \alpha_p )}_j (t) =
P^{(\alpha,\beta)}_j (t)$, we obtain the rather complicated result
\begin{eqnarray}
&&E^J_{jk} (\nu; \alpha, \beta | \zeta)
= 2^{2 \nu - 1} \, \frac{\Gamma (\nu)}{\Gamma (\frac{1}{2})}
\frac{\Gamma (j + \alpha + 1)}{\Gamma (j + \alpha + \beta + 1)}
\frac{\Gamma (k + \nu + \frac{1}{2})}{\Gamma (2k + 2\nu)}
\\
&&\quad\times \theta_{jk}\,
\sum_{\ell = 0}^{j - k}
(-1)^\ell
\frac{\Gamma (j + k + \ell + \alpha + \beta + 1)}{
\Gamma (\ell + 1)\Gamma (j - k - \ell + 1)\Gamma (k + \ell + \alpha + 1)}
(2 - \zeta)^{- \ell - k}
{_2F_1}
\left( \left. {- \ell, k + \nu + \frac{1}{2} \atop 2k + 2\nu + 1}
\right| \zeta \right) , \nonumber
\end{eqnarray}
which significantly simplifies for the particular values of the indices
$\alpha = \beta = \mu - \frac{1}{2}$ when Jacobi polynomials degenerate
into Gegenbauer ones. Therefore, taking the latter as a set of expansion
functions for our purposes, i.e. ${\cal P}^{( \alpha_p )}_j (t)
= C^\mu_j (t)$ we have
\begin{eqnarray}
\label{CoeffGegebauer}
&&E^G_{jk} (\nu; \mu | \zeta)
=
\frac{1}{2} \theta_{jk} \left[ 1 + (-1)^{j - k} \right]
\frac{\Gamma (\nu)}{\Gamma (\mu)}
\frac{ (-1)^{\frac{j - k}{2}}
\Gamma \left( \mu + \frac{j + k}{2} \right) }
{ \Gamma \left( \nu + k \right)
\Gamma \left( 1 + \frac{j - k}{2} \right) } \\
&&\hspace{8cm}\times ( 2 - \zeta )^{- k}
{_2F_1}
\left( \left. {- \frac{j - k}{2}, \mu + \frac{j + k}{2} \atop \nu + k + 1}
\right| \frac{\zeta^2}{(2 - \zeta)^2} \right). \nonumber
\end{eqnarray}
The fact that only even $j - k$-moments contribute is obvious since
the non-forward distributions are even functions of $\eta =
\frac{\zeta}{2 - \zeta}$ due to time-reversal invariance and hermiticity
\cite{Ji98}, --- as a result only even powers of $\eta$ can appear in
the expansion.

Now we are in a position to give results for the evolution operator,
${\mbox{\boldmath$\cal E$}}$, and ${\mbox{\boldmath$B$}}$-matrix
in two-loop approximation.

%%%%%%%%%%%%%%%%%%%%%%%%%%%%%%%%%%%%%%%%%%%%%%%%%%%%%%%%%%%%%%%%%%%%%
\subsection{Explicit NLO solution.}
%%%%%%%%%%%%%%%%%%%%%%%%%%%%%%%%%%%%%%%%%%%%%%%%%%%%%%%%%%%%%%%%%%%%%

In NLO the evolution operator satisfies the equation
\begin{equation}
\frac{d}{d\ln \alpha_s (Q^2)}
\mbox{\boldmath$\cal E$}_j \left( \alpha_s (Q^2), \alpha_s (Q_0^2) \right)
= - \frac{1}{\beta_0}
\left\{
\mbox{\boldmath$\gamma$}^{{\rm D}(0)}_j
+ \frac{\alpha_s (Q^2)}{2 \pi}
\mbox{\boldmath$R$}_j
\right\}
\mbox{\boldmath$\cal E$}_j \left( \alpha_s (Q^2), \alpha_s (Q_0^2) \right),
\end{equation}
where
\begin{equation}
\mbox{\boldmath$R$}_j =
\mbox{\boldmath$\gamma$}^{{\rm D}(1)}_j
- \frac{\beta_1}{2 \beta_0}
\mbox{\boldmath$\gamma$}^{{\rm D}(0)}_j ,
\end{equation}
and the boundary condition $\mbox{\boldmath$\cal E$}_j \left(
\alpha_s (Q_0^2), \alpha_s (Q_0^2) \right) = \1$ with the unit
matrix $\1 = \left( {1\,0\atop 0\,1} \right)$. Here $\beta_0$
and $\beta_1$ are the first and second coefficient in the
perturbative expansion of the QCD $\beta$-function $\frac{\beta}{\g}
= \frac{\alpha_s}{4\pi}\ \beta_0 + \left( \frac{\alpha_s}{4\pi}
\right)^2 \beta_1 + \dots$ and read $\beta_0 = \frac{4}{3} T_F N_f
- \frac{11}{3} C_A$ and $\beta_1 = \frac{10}{3} C_A N_f + 2 C_F N_f
- \frac{34}{3} C_A^2$, respectively.

The solution of the above equation is \cite{FurPet82,GluReyVog90}
\begin{eqnarray}
\mbox{\boldmath$\cal E$}_j \left( \alpha_s (Q^2), \alpha_s (Q_0^2) \right)
&=& \left(
\mbox{\boldmath$P$}^+_j
- \frac{\alpha_s (Q^2) - \alpha_s (Q^2_0)}{2 \pi}
\frac{1}{\beta_0}
\mbox{\boldmath$P$}^+_j
\mbox{\boldmath$R$}_j
\mbox{\boldmath$P$}^+_j
\right)
\left(
\frac{\alpha_s (Q^2_0)}{\alpha_s (Q^2)}
\right)^{\gamma^+_j / \beta_0} \nonumber\\
&-& \frac{\alpha_s (Q^2)}{2\pi}
\frac{
\mbox{\boldmath$P$}^-_j
\mbox{\boldmath$R$}_j
\mbox{\boldmath$P$}^+_j
}{\gamma^-_j - \gamma^+_j + \beta_0}
\left(
1 - \left( \frac{\alpha_s (Q^2_0)}{\alpha_s (Q^2)}
\right)^{\left( \gamma^-_j - \gamma^+_j + \beta_0 \right)/\beta_0}
\right)
\left(
\frac{\alpha_s (Q^2_0)}{\alpha_s (Q^2)}
\right)^{\gamma^+_j / \beta_0} \nonumber\\
&+& (+ \leftrightarrow -) ,
\end{eqnarray}
where we have introduced projection operators
\begin{equation}
\mbox{\boldmath$P$}^\pm_j
= \frac{\pm 1}{\gamma^+_j - \gamma^-_j}
\left(
\mbox{\boldmath$\gamma$}^{{\rm D}(0)}_j
- \gamma^\mp_j \1
\right) ,
\end{equation}
and the eigenvalues of the LO anomalous dimension matrix
\begin{equation}
\gamma^{\pm}_j
= \frac{1}{2}
\left(
{^{QQ}\!{\gamma}^{{\rm D} (0)}_j}
+
{^{GG}\!{\gamma}^{{\rm D} (0)}_j}
\pm
\sqrt{\left(
{^{QQ}\!{\gamma}^{{\rm D} (0)}_j}
-
{^{GG}\!{\gamma}^{{\rm D} (0)}_j}
\right)^2 + 4
{^{GQ}\!{\gamma}^{{\rm D} (0)}_j}
{^{QG}\!{\gamma}^{{\rm D} (0)}_j}
}
\right).
\end{equation}

The ${\mbox{\boldmath$B$}}$-matrix which fixes the corrections
to the eigenfunctions of the NLO ER-BL kernels is given by
\begin{equation}
{\mbox{\boldmath$B$}} = \1 + {\mbox{\boldmath$B$}}^{(1)},
\end{equation}
where ${\mbox{\boldmath$B$}}^{(1)}$ is determined by the first order
differential equation \cite{BelMue98b}
\begin{equation}
\frac{d}{d\ln \alpha_s (Q^2)}
{\mbox{\boldmath$B$}}^{(1)} \left( \alpha_s (Q^2), \alpha_s (Q_0^2) \right)
= - \frac{1}{\beta_0}
\left\{
\left[
\mbox{\boldmath$\gamma$}^{{\rm D}(0)} ,
{\mbox{\boldmath$B$}}^{(1)} \left( \alpha_s (Q^2), \alpha_s (Q_0^2) \right)
\right]_-
+ \frac{\alpha_s (Q^2)}{2 \pi}
{\mbox{\boldmath$\gamma$}}^{{\rm ND}(1)}
\right\} ,
\end{equation}
and reads ($j > k$)
\begin{eqnarray}
{\mbox{\boldmath$B$}}^{(1)}_{jk}
&=& - \frac{\alpha_s (Q^2)}{2\pi}
\left(
\frac{
\mbox{\boldmath$P$}^+_j
\mbox{\boldmath$\gamma$}^{{\rm ND}(1)}_{jk}
\mbox{\boldmath$P$}^+_k
}{\gamma^+_j - \gamma^+_k + \beta_0}
\left(
1 - \left( \frac{\alpha_s (Q^2_0)}{\alpha_s (Q^2)}
\right)^{\left( \gamma^+_j - \gamma^+_k + \beta_0 \right)/\beta_0}
\right) \right. \nonumber\\
&&\qquad\qquad +
\left.\frac{
\mbox{\boldmath$P$}^+_j
\mbox{\boldmath$\gamma$}^{{\rm ND}(1)}_{jk}
\mbox{\boldmath$P$}^-_k
}{\gamma^+_j - \gamma^-_k + \beta_0}
\left(
1 - \left( \frac{\alpha_s (Q^2_0)}{\alpha_s (Q^2)}
\right)^{\left( \gamma^+_j - \gamma^-_k + \beta_0 \right)/\beta_0}
\right) \right) \nonumber\\
&&\qquad\qquad +\ (+ \leftrightarrow -).
\end{eqnarray}

The two-loop forward anomalous dimension matrix in the singlet sector
has been evaluated in Refs.\ \cite{Singlet80,FloKouLac81}.
The non-diagonal entries, $\mbox{\boldmath$\gamma$}^{{\rm ND}(1)}$,
of the full anomalous dimension matrix $\mbox{\boldmath$\gamma$} =
\mbox{\boldmath$\gamma$}^{{\rm D}} + \mbox{\boldmath$\gamma$}^{{\rm ND}}$
of the conformal operators have become available quite recently
\cite{BelMue98a,BelMue98b}. For the reader's convenience both
of the above ingredients are summarized in appendices \ref{SingletADs}
and \ref{NonDiagAD}, respectively.

Finally, the coupling constant in the corresponding order is given by the
following inverse-log expansion
\begin{equation}
\alpha_s(Q^2) = - \frac{4\pi}{\beta_0
\ln ( Q^2 / \Lambda^2_{\overline{\rm MS}})}
\left(
1 + \frac{\beta_1}{\beta_0^2}
\frac{\ln \, \ln ( Q^2 / \Lambda^2_{\overline{\rm MS}} )}{
\ln ( Q^2 / \Lambda^2_{\overline{\rm MS}}) }
\right),
\end{equation}
Note, however, that due to the fact that the input scale, $Q_0$, ---
at which the initial conditions considered below are defined --- is
very low it might be more reliable and accurate to obtain $\alpha_s$
in two-loop approximation by solving the exact transcendental equation
\begin{equation}
- \beta_0 \ln \frac{Q^2}{\Lambda^2_{\overline{\rm MS}}}
= \frac{4 \pi}{\alpha_s (Q^2)}
- \frac{\beta_1}{\beta_0}
\ln \left( - \frac{4 \pi}{\beta_0 \alpha_s (Q^2)}
- \frac{\beta_1}{\beta_0^2} \right).
\end{equation}
This leads to $\sim 15\%$ discrepancy between them for $Q^2 = 0.6\
{\rm GeV}^2$ which goes down to $3\%$ for $Q^2 = 4\ {\rm GeV}^2$.
However, we leave aside this source of theoretical uncertainty, as
being conceptually irrelevant for our study it presents only an extra
source of unnecessary complication.

%%%%%%%%%%%%%%%%%%%%%%%%%%%%%%%%%%%%%%%%%%%%%%%%%%%%%%%%%%%%%%%%%%%%%
\section{Evolution of the model distributions.}
%%%%%%%%%%%%%%%%%%%%%%%%%%%%%%%%%%%%%%%%%%%%%%%%%%%%%%%%%%%%%%%%%%%%%

For practical applications we have chosen the indices of the Jacobi
polynomials in the simplest way, i.e.\ $\alpha = \beta = 0$. In this
case the former coincide with Legendre polynomials, $P_j (t)
= P^{(0,0)}_j (t) = C^{1/2}_j (t)$ with $w (t) = 1$ and $n_j =
(2j + 1)/2$ in Eqs.\ (\ref{PolySeries},\ref{PolyMatrix}). From the
last equality we can read off explicit expressions for the expansion
coefficients $E_{jk}$ in Eq.\ (\ref{CoeffGegebauer}). Contrary to our
previous procedure advocated in Ref.\ \cite{BelGeyMueSch97} where the
evolution of a limited number of points in the discrete $\{x,\zeta\}$-plane
has been performed exactly and interpolated afterwards by a smooth
functions in spline approximation, presently we evolve the expansion
coefficients of the non-forward parton distribution function in the
series over orthogonal polynomials and then sum them back. Therefore,
no smoothing recipes are required but the result depends on the
truncation parameter $N_{\rm max}$. To achieve rather high
reconstruction accuracy, i.e.\ $\sim 10^{-2}-10^{-3}$, we have retained
up to $N_{\rm max} = 100$ polynomials in the series\footnote{The
$C\!\!+\!\!+$ code used in the calculation is available via
{\sf http://www.physik.uni-regensburg.de/\~{}nil17791}.}
(\ref{PolySeries}). Even better precision, $10^{-4}$, is feasible
by doubling the number, $N_{\rm max} = 100$, of polynomials in the
expansion at the price of the much more time consuming procedure.
Note that only the region around the crossover point $x = \zeta$ is
sensitive to $N_{\rm max}$ due to rather rapid change of the curve
on the small interval of $x$. Taking only $N_{\rm max} \sim 50$ the
reconstructions accuracy decreases down to $10^{-1}-10^{-2}$.

Note that due to symmetry properties of the
distribution functions only those moments survive which acquire
an operator content and evolve with anomalous dimensions deduced
from OPE. This is in contrast to the case of distributions,
${^A\!{\cal O}}$, with the support $0 \leq x \leq 1$ where all
moments enter on equal grounds due to lack of charge conjugation
symmetry in the parametrization of the hadronic matrix elements in Eqs.\
(\ref{NFPD-Q},\ref{NFPD-G},\ref{Q-decomp},\ref{G-decomp}). They
were considered in Ref.\ \cite{BelMueNidSch98} and it was crucial
there to use the analytically continued anomalous
dimensions\footnote{In the above mentioned paper we have
expanded the anomalous dimensions in the series $\gamma_j = \sum_{l,\, m}
c_{lm} \frac{\ln^l(j + 1)}{(j + 1)^m}$ with the first few terms kept
in the expansion which provides a highly accurate approximation,
sufficient for numerical studies.} (see appendix \ref{SingletADs}) since
otherwise the convergence of the series will break down already for
$N_{\rm max} \sim 10$. This is so in spite of the fact that the relative
difference between the former and the ones deduced from an OPE analysis
with $\sigma = (-1)^{j + 1}$ (see appendix \ref{SingletADs}) left intact
is negligible ($\leq 10^{-4}$). The reason is the factorial growth of
the expansion coefficients which weight the contribution of moments.

Below we report on our study of the $Q^2$-dependence for the
distribution functions proposed in Ref.\ \cite{Rad98}. We will,
however, neither speculate on the physical relevance of the latter nor
discuss their adequacy to the real world since this is a disputable issue
in the lack of any experimental data, but we rather accept them in order
to test our formalism. We have chosen for our analysis $N_f = 4$ and
$\Lambda^{(4)}_{\overline{\rm MS}} = 246\ {\rm MeV}$.

%%%%%%%%%%%%%%%%%%%%%%%%%%%%%%%%%%%%%%%%%%%%%%%%%%%%%%%%%%%%%%%%%%%%%
%                         Figure 2
%%%%%%%%%%%%%%%%%%%%%%%%%%%%%%%%%%%%%%%%%%%%%%%%%%%%%%%%%%%%%%%%%%%%%
\begin{figure}
\begin{center}
\vspace{6.5cm}
\hspace{0cm}
\mbox{
\unitlength1mm
\begin{picture}(150,110)(0,0)
\put(17,126){\inclfig{11.84}{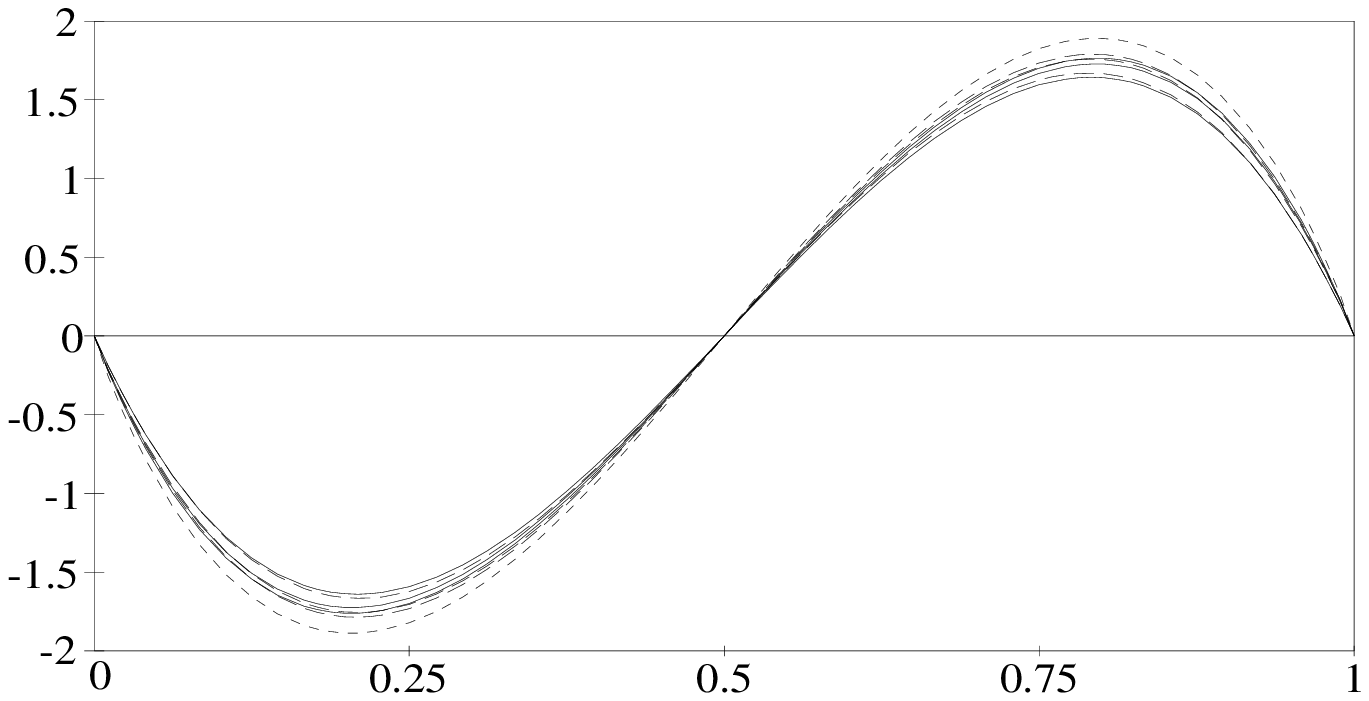}}
\put(125,177){(a)}
\put(35,177){$\zeta = 1.0$}
\put(121,125){$x$}
\put(10,150){
\rotate{$\scriptstyle {\cal Q} \left(x,\zeta,Q^2\right)$}}
\put(20,58){\inclfig{11.5}{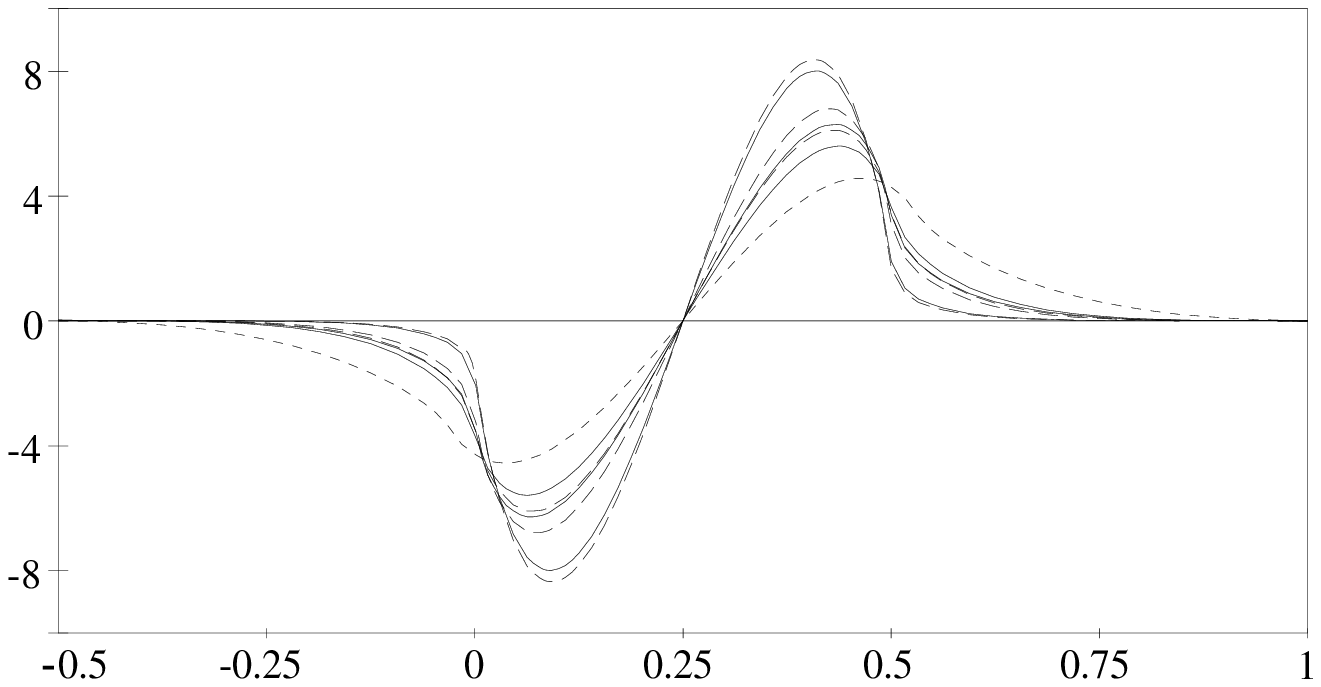}}
\put(35,109){$\zeta = 0.5$}
\put(125,109){(b)}
\put(121,57){$x$}
\put(10,82){
\rotate{$\scriptstyle {\cal Q} \left(x,\zeta,Q^2\right)$}}
\put(15.5,-10){\inclfig{12.2}{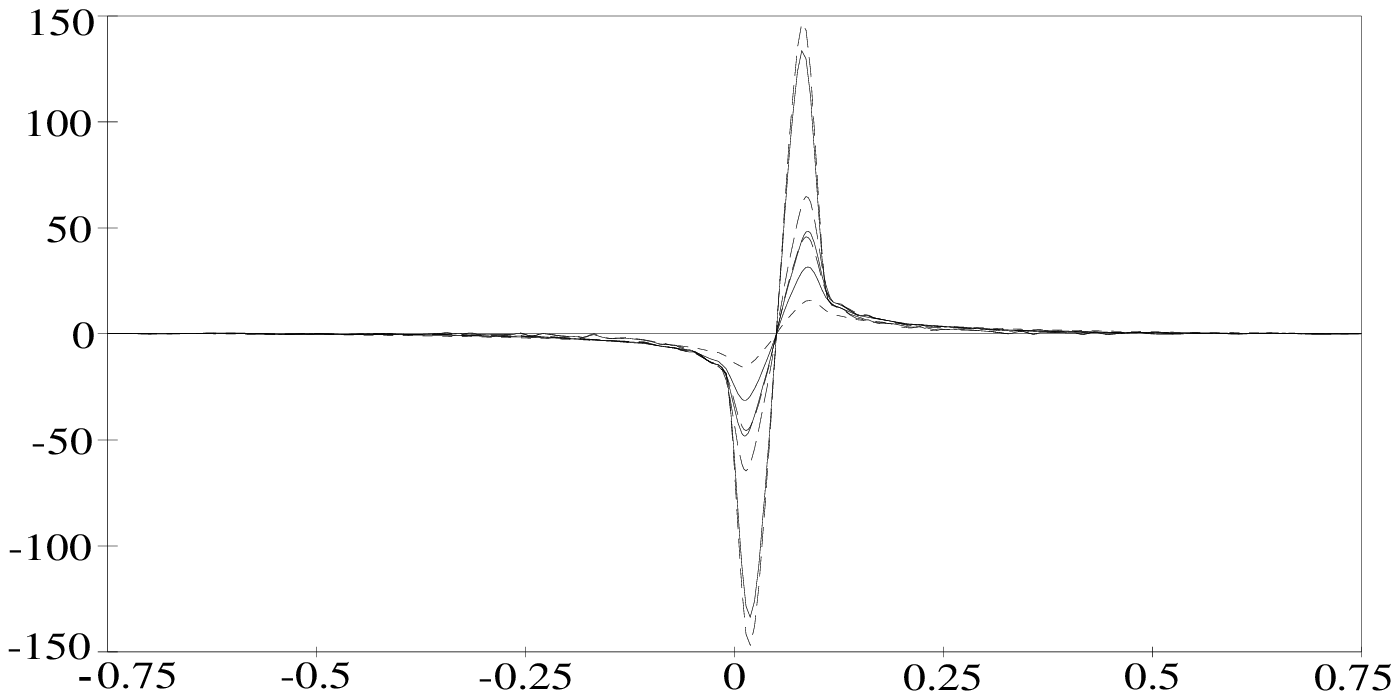}}
\put(125,41){(c)}
\put(35,41){$\zeta = 0.1$}
\put(121,-12){$x$}
\put(10,13){
\rotate{$\scriptstyle {\cal Q} \left(x,\zeta,Q^2\right)$}}
\end{picture}
}
\vspace{1cm}
\end{center}
\caption{\label{QuarkDistr} Evolution of the non-forward singlet quark
distributions ${\cal Q} (x, \zeta)$. The input function at $Q_0 = 0.7\
{\rm GeV}$ is shown by the short-dashed line at different $\zeta$'s. The
full curves moving away from the initial function correspond to
LO results for $Q^2 = 10,\ 10^2,\ 10^{14}\ {\rm GeV}^2$, respectively.
The long-dashed lines give the NLO results for the same values of the
momentum scale in the same order.}
\end{figure}

%%%%%%%%%%%%%%%%%%%%%%%%%%%%%%%%%%%%%%%%%%%%%%%%%%%%%%%%%%%%%%%%%%%%%
%                         Figure 3
%%%%%%%%%%%%%%%%%%%%%%%%%%%%%%%%%%%%%%%%%%%%%%%%%%%%%%%%%%%%%%%%%%%%%
\begin{figure}
\begin{center}
\vspace{6.5cm}
\hspace{0cm}
\mbox{
\unitlength1mm
\begin{picture}(150,110)(0,0)
\put(18,127){\inclfig{11.73}{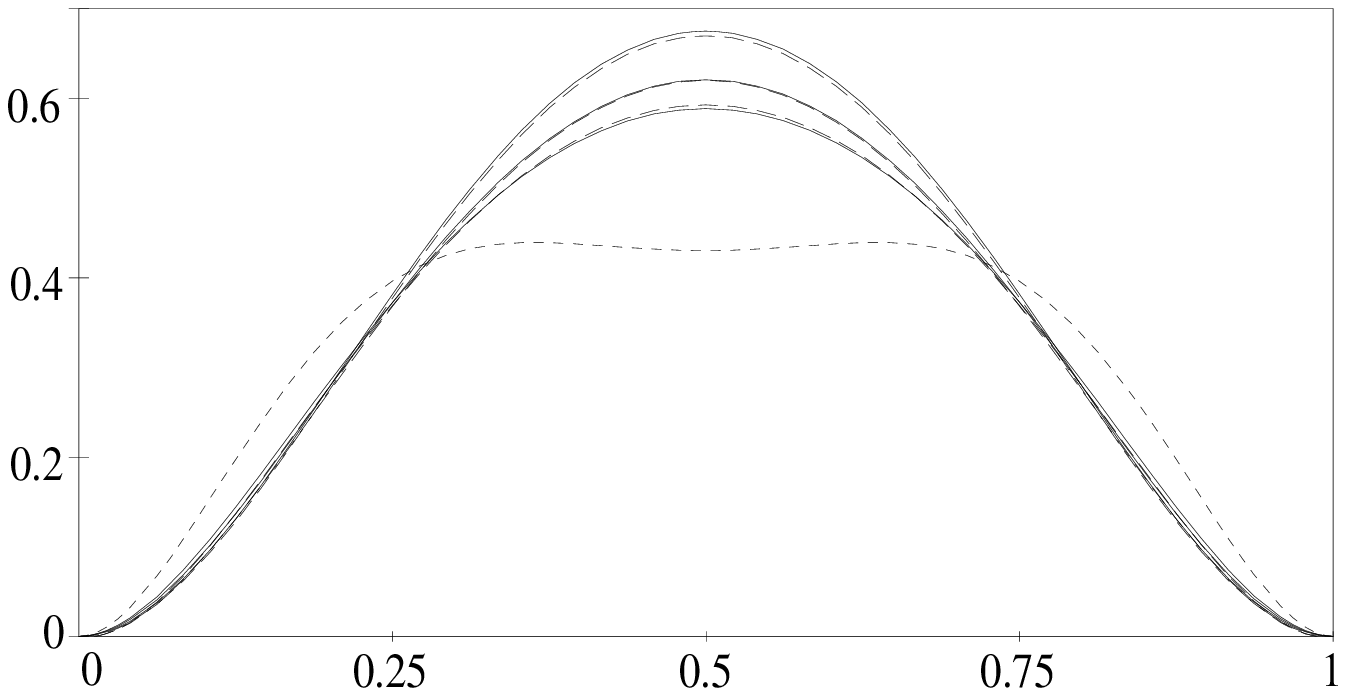}}
\put(121,177){(a)}
\put(35,177){$\zeta = 1.0$}
\put(121,125){$x$}
\put(10,150){
\rotate{$\scriptstyle {\cal G} \left(x,\zeta,Q^2\right)$}}
\put(16,60){\inclfig{11.9}{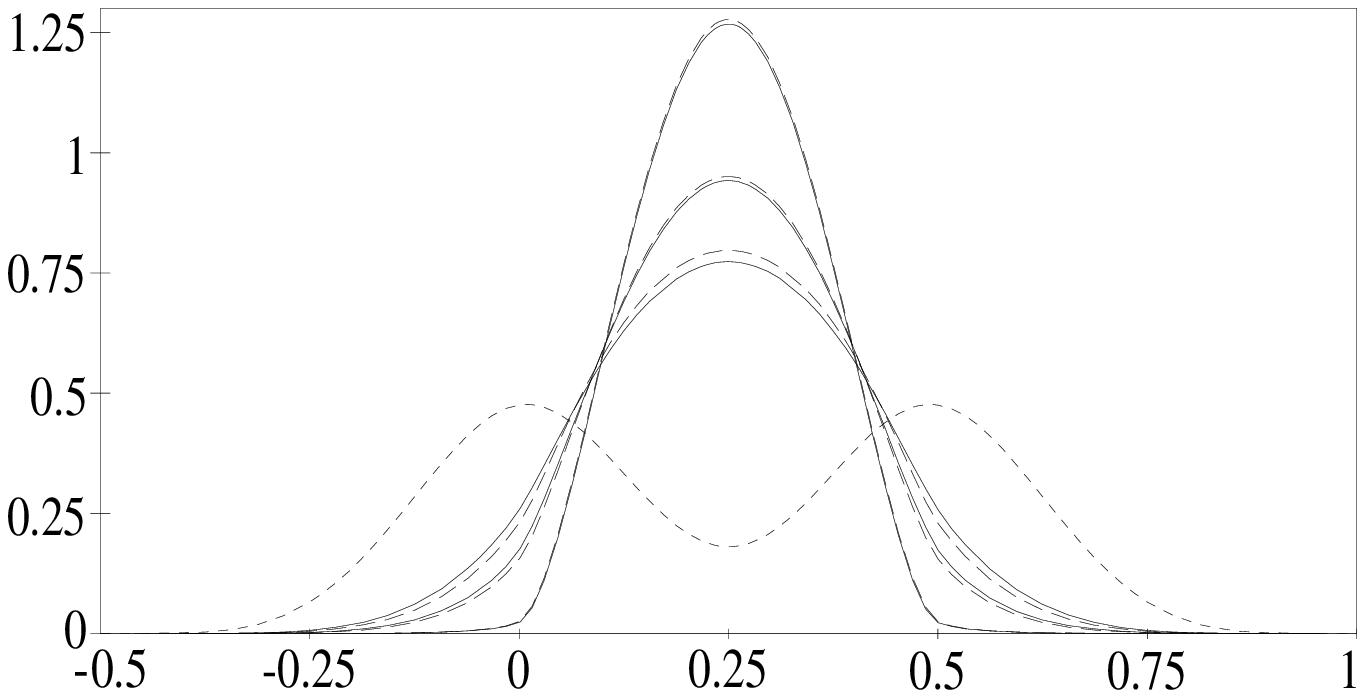}}
\put(35,109){$\zeta = 0.5$}
\put(121,109){(b)}
\put(121,57){$x$}
\put(10,82){
\rotate{$\scriptstyle {\cal G} \left(x,\zeta,Q^2\right)$}}
\put(21,-9){\inclfig{11.58}{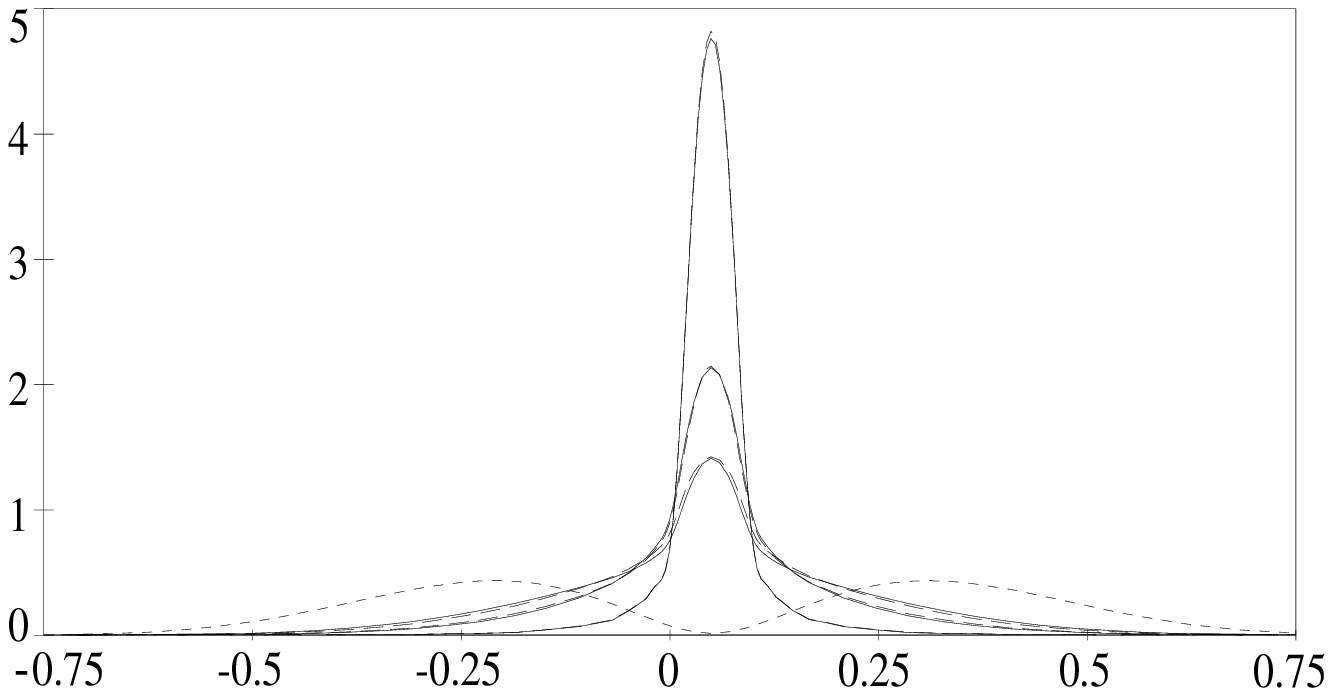}}
\put(121,41){(c)}
\put(35,41){$\zeta = 0.1$}
\put(121,-12){$x$}
\put(10,13){
\rotate{$\scriptstyle {\cal G} \left(x,\zeta,Q^2\right)$}}
\end{picture}
}
\vspace{1cm}
\end{center}
\caption{\label{GluonDistr} Same as Fig.\ \protect\ref{QuarkDistr}
but for the gluon non-forward distribution ${\cal G} (x, \zeta)$. The
conventions are the same as in Fig.\ \protect\ref{QuarkDistr}.}
\end{figure}

The non-forward functions ${^A\!{\cal O}} (x, \zeta)$ in Eqs.\
(\ref{Q-decomp},\ref{G-decomp}) for the parton species
$A = Q,\,\bar Q,\, G$ are defined in terms of the double
distribution function ${^A\!F}(y, z)$ via the following relation
\cite{Rad97}
\begin{equation}
\label{nontodouble}
{^A\!{\cal O}} (x, \zeta, Q^2_0)
=  \int_0^1 d y \int_0^1 d z \, {^A\!F} (y, z, Q^2_0) \,
\theta (1 - y - z) \delta (x - y - \zeta z).
\end{equation}
The functional dependence of $^A\!F$ in the $y$-subspace is given
by the shape of the forward parton density while its
$\frac{z}{\bar y}$-dependence has to be similar to that of the
distribution amplitude. This results in the following model for
${^A\!F} (y, z, Q^2_0)$ \cite{Rad98}
\begin{equation}
\label{DDfunction}
{^A\!F} (y, z, Q^2_0) = A (y, Q^2_0) {^A\!\pi} (y, z),
\end{equation}
with the plausible profiles ${^A\!\pi} (y, z)$ \cite{Rad98}
\begin{equation}
\label{realprofile}
{^Q\!\pi} (y, z) = 6 \frac{z}{\bar y^3} ( \bar y - z ) ,\qquad
{^G\!\pi} (y, z) = 30 \frac{z^2}{\bar y^5} ( \bar y - z )^2 ,
\end{equation}
for quarks and gluons, respectively. In Eq.\ (\ref{DDfunction}) the
function $A (y, Q^2_0) = q (y),\ \bar q (y),\ y g (y)$ is an ordinary
forward parton density measured in deep inelastic scattering
at a low normalization point. We use the CTEQ4LQ parametrization
\cite{Lai97} of the parton densities at the momentum scale
$Q_0 = 0.7\ {\rm GeV}$:
\begin{eqnarray}
x u_v (x) = x ( u - \bar u ) (x)
&=& 1.315 x^{0.573} (1 - x)^{3.281} (1 + 10.614 x^{1.034}),\nonumber\\
x d_v (x) = x ( d - \bar d ) (x)
&=& 0.852 x^{0.573} (1 - x)^{4.060} (1 + 4.852 x^{0.693}) ,\nonumber\\
x ( \bar u + \bar d ) (x)
&=& 0.578 x^{0.143} (1 - x)^{7.293} (1 + 1.858 x^{1.000}) ,\\
x g (x)
&=& 39.873 x^{1.889} (1 - x)^{5.389} (1 + 0.618 x^{0.474}) . \nonumber
\end{eqnarray}

The results of the evolution are shown in Figs.\ \ref{QuarkDistr},
\ref{GluonDistr} for singlet quark and gluon distributions, respectively.
Due to the low input momentum scale, $Q_0$, the change of the shape
is very prominent already for $Q^2 = 10\ {\rm GeV}^2$. We have plotted
the curves obtained using the LO (full lines) and NLO (dashed
lines) formulae. The difference between them is especially sizable
($\sim 10-30\%$) for quarks at small $\zeta$'s and moderately large
$Q^2$. The limiting lines correspond to the extremely large momentum
scales, $Q^2 = 10^{14}\ {\rm GeV}^2$, when the distributions
approximately reach their asymptotic forms ($Q^2 \to \infty$) \cite{Rad98}:
\begin{equation}
{\cal Q}_{\rm as} (x, \zeta)
\propto \frac{x}{\zeta^2} \left( 1 - \frac{x}{\zeta} \right)
\left( 2 \frac{x}{\zeta} - 1 \right) \theta (\zeta - x),\qquad
{\cal G}_{\rm as} (x, \zeta)
\propto \frac{x^2}{\zeta^3} \left( 1 - \frac{x^2}{\zeta^2} \right)
\theta (\zeta - x) .
\end{equation}

%%%%%%%%%%%%%%%%%%%%%%%%%%%%%%%%%%%%%%%%%%%%%%%%%%%%%%%%%%%%%%%%%%%%%
\section{Conclusions.}
%%%%%%%%%%%%%%%%%%%%%%%%%%%%%%%%%%%%%%%%%%%%%%%%%%%%%%%%%%%%%%%%%%%%%

In the present paper we have generalized the formalism for the
solution of NLO evolution equations for the non-forward parton
distributions developed by us previously for the non-singlet case
\cite{BelMueNidSch98,BelGeyMueSch97} to the flavour singlet channel.
It is based on the use of the Jacobi polynomials to reconstruct the
$(x,\zeta)$-dependence of the function from the multiplicatively
renormalizable conformal moments which diagonalize the generalized
ER-BL evolution equation. Due to relatively fast convergence of the
series in orthogonal polynomials we limit ourselves to at most
$N_{\rm max} \leq 100$ terms. The resummation can still
be handled numerically and the precision achieved is sufficiently
high for any practical application. Let us note that in absence of
analytic expressions for the complete NLO non-forward kernels in the
singlet channel (at least for the time being) there is no feasible
alternative to the procedure described above.

We did not analyze in full, however, how the choice for the parameters
$\alpha$ and $\beta$ influences the convergence properties of the
series. One should note that a clever choice can lead to an increase
of reconstruction accuracy by an order of magnitude for the same
$N_{\rm max}$ \cite{Kri87}.

We have studied the general pattern of evolution obeyed by the models
for the non-forward distributions proposed recently. We have found that
the difference between LO and NLO does not exceed $10-30\%$ which
suggests that the latter effects could be taken into account
provided very high accuracy data points will be available which
seems to be a very hard experimental task.

An application of our considerations to the evolution of the
parity-odd densities is straightforward since the non-diagonal elements,
$\mbox{\boldmath$\gamma$}^{{\rm ND}}$, of the corresponding anomalous
dimensions are known from Ref.\ \cite{BelMue98b} while the analytically
continued two-loop forward entities \cite{MerNeer96} were derived in
\cite{GRCV95}.

\vspace{0.5cm}

We acknowledge helpful discussions with A.V. Radyushkin and
would like to thank him for reading of the manuscript and useful
comments. This work was supported by BMBF and the Alexander von
Humboldt Foundation (A.B.).

\appendix

\setcounter{section}{0}
\setcounter{equation}{0}
\renewcommand{\theequation}{\Alph{section}.\arabic{equation}}

%%%%%%%%%%%%%%%%%%%%%%%%%%%%%%%%%%%%%%%%%%%%%%%%%%%%%%%%%%%%%%%%%%%%%
\section{Diagonal (forward) anomalous dimensions.}
\label{SingletADs}
%%%%%%%%%%%%%%%%%%%%%%%%%%%%%%%%%%%%%%%%%%%%%%%%%%%%%%%%%%%%%%%%%%%%%

Let us add few remarks concerning the NLO anomalous dimensions for
DIS. It should be noted that beyond leading order the moments of
the DGLAP splitting kernels coincide with the anomalous dimensions
available in the literature evaluated by OPE methods only for even
(odd) moments provided we are interested in the crossing odd (even)
combination of quark--anti-quark species. We give below analytically
continued anomalous dimensions from even (odd) to all values of moments
$j$ \cite{CurFurPet80,GluReyVog90} where the above unfortunate feature
is naturally withdrawn.

The one-loop quantities are well known and read \cite{GroWil74}
\begin{eqnarray}
\label{LOAD}
{^{QQ}\!\gamma}_{j}^{{\rm fw}(0)}
&=&
- C_F \left(
- 4 \psi( j + 2 ) + 4 \psi(1) + \frac{2}{( j + 1 )( j + 2 )} + 3
\right) , \\
{^{QG}\!\gamma}_{j}^{{\rm fw}(0)}
&=& - 4 N_f T_F
\frac{j^2 + 3 j + 4}{( j + 1 )( j + 2 )( j + 3 )} , \\
{^{GQ}\!\gamma}_{j}^{{\rm fw}(0)}
&=& - 2 C_F
\frac{j^2 + 3 j + 4}{j( j + 1 )( j + 2 )} , \\
\label{last-AD}
{^{GG}\!\gamma}_{j}^{{\rm fw}(0)}
&=&
- C_A \left(
- 4 \psi( j + 2 ) + 4 \psi(1)
+ 8 \frac{j^2 + 3 j + 3}{j( j + 1 )( j + 2 )( j + 3 )}
- \frac{\beta_0}{C_A}
\right) .
\end{eqnarray}

The anomalous dimensions at ${\cal O}(\alpha_s^2)$ are
given by \cite{FloKouLac81}:
\begin{eqnarray}
{^{QQ}\!\gamma^{{\rm fw}\,{\rm NS}(1)}_j} (\sigma)
&=&
\left( C_F^2 - \frac{1}{2} C_F C_A \right)
\left\{
\frac{4 (2 j + 3)}{(j + 1)^2 (j + 2)^2} S(j + 1)
- 2 \frac{3 j^3 + 10 j^2 + 11 j + 3}{(j + 1)^3 (j + 2)^3}
\right.
\nonumber\\
&+& 4 \left( 2 S_1(j + 1) - \frac{1}{(j + 1)(j + 2)} \right)
\left(
S_2(j + 1) - S_2^{\prime} (j + 1)
\right) \nonumber\\
&+& \left.
16 \tilde{S}(j + 1) + 6 S_2(j + 1)
- \frac{3}{4} - 2 S_3^{\prime} (j + 1)
+ 4 \sigma
\frac {2 j^2 + 6 j + 5}{(j + 1)^3 (j + 2)^3} \right\} \nonumber\\
&+& C_F C_A
\left\{ S_1(j + 1)
\left( \frac{134}{9} + \frac{2 (2 j + 3)}{(j + 1)^2 (j + 2)^2} \right)
\right. \nonumber\\
&-& 4 S_1 (j + 1) S_2(j + 1)
+ S_2(j + 1) \left( - \frac{13}{3} + \frac{2}{(j + 1) (j + 2)} \right)
\nonumber\\
&-&
\left.
\frac{43}{24} -
\frac{1}{9} \frac{151 j^4 + 867 j^3 + 1792 j^2 + 1590 j + 523
}{(j + 1)^3 (j + 2)^3} \right\} \nonumber\\
&+&
C_F T_F N_f
\left\{
- \frac{40}{9} S_1(j + 1) + \frac{8}{3} S_2(j + 1) + \frac{1}{3}
+ \frac{4}{9} \frac{ 11 j^2 + 27 j + 13}{(j + 1)^2 (j + 2)^2}
\right\} ,\\
&& \nonumber\\
{^{QQ}\!\gamma^{{\rm fw}(1)}_j} &=&
{^{QQ}\!\gamma^{{\rm fw}\,{\rm NS}(1)}_j} (\sigma = 1)
- 4 C_F T_F N_f
\frac{5 j^5 + 57 j^4 + 227 j^3 + 427 j^2 + 404 j + 160
}{j(j + 1)^3 (j + 2)^3 (j + 3)^2} ,\\
&& \nonumber\\
{^{QG}\!\gamma^{{\rm fw}(1)}_j}
&=&
- 2 C_A T_F N_f \left\{
\left( -2 S_1^2 (j + 1) + 2 S_2(j + 1)
- 2 S_2^{\prime} (j + 1) \right)
\frac{j^2 + 3 j + 4}{(j + 1) (j + 2) (j + 3)} \right. \nonumber\\
&+&
2 \frac{j^9 + 15 j^8 + 99 j^7 + 382 j^6 + 963 j^5 + 1711 j^4
+ 2273 j^3 + 2252 j^2 + 1488 j + 480}{j (j + 1)^3 (j + 2)^3 (j + 3)^3}
\nonumber\\
&+&
\left.
8 \frac{(2 j + 5)}{(j + 2)^2 (j + 3)^2} S_1 (j + 1)
\right\} \nonumber\\
&-&
2 C_F T_F N_f
\left\{ \left( 2 S_1^2(j + 1) - 2 S_2(j + 1) + 5 \right)
\frac{j^2 + 3 j + 4}{(j + 1)(j + 2)(j + 3)} \right. \nonumber\\
&-& \left.
4 \frac{S_1(j + 1)}{(j + 1)^2}
+ \frac{11 j^4 + 70 j^3 + 159 j^2 + 160 j + 64}{(j + 1)^3(j + 2)^3(j + 3)}
\right\} ,\\
&& \nonumber\\
{^{GQ}\gamma^{{\rm fw}(1)}_j} &=&
- C_F^2 \left\{
\left( -2 S_1^2(j + 1) + 10 S_1(j + 1) - 2 S_2(j + 1) \right)
\frac{j^2 + 3 j + 4}{j(j + 1)(j + 2)} \right. \nonumber\\
&-&
\left. 4 \frac{S_1(j + 1)}{(j + 2)^2}
- \frac{12 j^6 + 102 j^5 + 373 j^4 + 740 j^3 + 821 j^2 + 464 j + 96
}{j(j + 1)^3(j + 2)^3} \right\} \nonumber\\
&-& 2 C_A C_F \left\{
\left( S_1^2(j + 1) + S_2(j + 1) - S_2^{\prime}(j + 1) \right)
\frac{j^2 + 3 j + 4}{j(j + 1)(j + 2)}\right. \nonumber\\
&+&
\frac{1}{9}
\frac {109 j^9 + 1602 j^8 + 10292 j^7 + 38022 j^6 + 88673 j^5
+ 133818 j^4 + 128014 j^3}{j^2(j + 1)^3(j + 2)^3(j + 3)^2} \nonumber\\
&+&
\left.
\frac{1}{9}
\frac {72582 j^2 + 21384 j + 2592}{j^2(j + 1)^3(j + 2)^3(j + 3)^2}
- \frac{1}{3} \frac{17 j^4 + 68 j^3 + 143 j^2 + 128 j + 24
}{j^2(j + 1)^2 (j + 2)} S_1(j + 1) \right\} \nonumber\\
&-&
\frac{8}{3} C_F T_F N_f
\left\{ \left( S_1(j + 1)- \frac{8}{3} \right)
\frac{j^2 + 3 j + 4}{j(j + 1)(j + 2)} + \frac{1}{(j + 2)^2} \right\} ,\\
&& \nonumber\\
{^{GG}\gamma^{{\rm fw}(1)}_j}
&=&
C_A T_F N_f \left\{
-\frac{40}{9} S_1(j + 1) + \frac{8}{3} + \frac{8}{9}
\frac{19 j^4 + 114 j^3 + 275 j^2 + 312 j + 138}{j(j + 1)^2(j + 2)^2(j + 3)}
\right\} \nonumber\\
&+& C_F T_F N_f
\left\{ 2
+ 4 \frac{2 j^6 + 16 j^5 + 51 j^4 + 74 j^3 + 41 j^2 - 8 j - 16
}{j(j + 1)^3(j + 2)^3(j + 3)} \right\} \nonumber\\
&+& C_A^2
\left\{ \frac{134}{9} S_1(j + 1) + 16 S_1(j + 1)
\frac{2 j^5 + 15 j^4 + 48 j^3 + 81 j^2 + 66 j + 18
}{ j^2(j + 1)^2(j + 2)^2(j + 3)^2} \right. \nonumber\\
&-& \frac{16}{3} + 8 S_2^{\prime}(j + 1)
\frac{j^2 + 3 j + 3}{j(j + 1)(j + 2)(j + 3)}
- 4 S_1(j + 1)S_2^{\prime}(j + 1) \nonumber\\
&+& 8 \tilde{S}(j + 1) - S_3^{\prime}(j + 1)
-\frac{1}{9}
\frac{457 j^9 + 6855 j^8 + 44428 j^7 + 163542 j^6
}{j^2(j + 1)^3(j + 2)^3(j + 3)^3} \nonumber\\
&-&
\left. \frac{1}{9}
\frac{376129 j^5 + 557883 j^4 + 529962 j^3 + 308808 j^2 + 101088 j + 15552
}{j^2(j + 1)^3(j + 2)^3(j + 3)^3} \right\} .
\end{eqnarray}

Here we should use the following expressions for the analytically
continued functions \cite{GluReyVog90}
\begin{eqnarray}
S_1 (j) &=& \gamma_E + \psi(j + 1),\\
S_2 (j) &=& \zeta (2) - \psi^\prime (j + 1),\\
S_3 (j) &=& \zeta (3) + \frac{1}{2} \psi^{\prime\prime} (j + 1),\\
S^\prime_\ell (j) &=& \frac{1}{2} (1 + \sigma) S_\ell \left(\frac{j}{2}\right)
+ \frac{1}{2} (1 - \sigma) S_\ell \left( \frac{j-1}{2} \right),\\
\widetilde{S} (j) &=& - \frac{5}{8} \zeta(3)
+ \sigma \left\{ \frac{S_1 (j)}{j^2}
- \frac{\zeta(2)}{2}\left(
\psi\left( \frac{j+1}{2} \right) - \psi \left( \frac{j}{2} \right) \right)
+ \int_0^1 dx x^{j-1} \frac{\mbox{Li}_2(x)}{1 + x} \right\},
\end{eqnarray}
where $\psi^{(\ell - 1)} (j) = \frac{d^{(\ell)} \Gamma(j)}{dj^{\ell}}$
is the poly-gamma function and $\gamma_E = - \psi (1)$ is the
Euler-Mascheroni constant.

The integral over the dilogarithm $\mbox{Li}_2(x)$ can be evaluated
in terms of an analytical function of $j$ by using the least square
fit approximation
\cite{GluReyVog90}:
\begin{equation}
\frac{\mbox{Li}_2 (x)}{1 + x}
\simeq
0.0030 + 1.0990 x -1.5463 x^2 + 3.2860 x^3 - 3.7887 x^4 + 1.7646 x^5 .
\end{equation}
The $\sigma = (-1)^{j + 1}$ takes the following values for ``non-singlet''
--- charge conjugation odd, $q_v$, and flavour NS, $q^{\rm NS}$, ---
combinations:
\begin{eqnarray}
\sigma = - 1 \quad\mbox{for}\quad
q_v = q - \bar q ,\qquad
\sigma = 1 \quad\mbox{for}\quad
q^{\rm NS} = ( u + \bar u ) - ( d + \bar d ) ,\qquad
\mbox{etc.,}
\end{eqnarray}
while the singlet distributions are evolved with $\sigma = 1$.

\setcounter{equation}{0}
\renewcommand{\theequation}{\Alph{section}.\arabic{equation}}

%%%%%%%%%%%%%%%%%%%%%%%%%%%%%%%%%%%%%%%%%%%%%%%%%%%%%%%%%%%%%%%%%%%%%
\section{Non-diagonal anomalous dimensions.}
\label{NonDiagAD}
%%%%%%%%%%%%%%%%%%%%%%%%%%%%%%%%%%%%%%%%%%%%%%%%%%%%%%%%%%%%%%%%%%%%%

The elements of the matrix $\mbox{\boldmath$\gamma$}^{{\rm ND}(1)}_{jk}$
derived in Ref.\ \cite{BelMue98b} read
\begin{eqnarray}
\label{andimND-QQ}
{^{QQ}\!\gamma}_{jk}^{{\rm ND}(1)}
&=&
\left(
{^{QQ}\!\gamma}_{j}^{{\rm D}(0)} - {^{QQ}\!\gamma}_{k}^{{\rm D}(0)}
\right)
\left\{
d_{jk}
\left(
\beta_0 - {^{QQ}\!\gamma}_{k}^{{\rm D}(0)}
\right)
+ {^{QQ}\!g}_{jk}
\right\} \\
&&\qquad\qquad\qquad\qquad\quad -
\left(
{^{QG}\!\gamma}_{j}^{{\rm D}(0)} - {^{QG}\!\gamma}_{k}^{{\rm D}(0)}
\right) d_{jk}
{^{GQ}\!\gamma}_{k}^{{\rm D}(0)}
+ {^{QG}\!\gamma}_{j}^{{\rm D}(0)} {^{GQ}\!g}_{jk}, \nonumber\\
\label{andimND-QG}
{^{QG}\!\gamma}_{jk}^{{\rm ND}(1)}
&=&
\left(
{^{QG}\!\gamma}_{j}^{{\rm D}(0)} - {^{QG}\!\gamma}_{k}^{{\rm D}(0)}
\right)
d_{jk}
\left( \beta_0 - {^{GG}\!\gamma}_{k}^{{\rm D}(0)} \right)
- \left(
{^{QQ}\!\gamma}_{j}^{{\rm D}(0)} - {^{QQ}\!\gamma}_{k}^{{\rm D}(0)}
\right)
d_{jk} {^{QG}\!\gamma}_{k}^{{\rm D}(0)} \\
&&\qquad\qquad\qquad\qquad\quad +
{^{QG}\!\gamma}_{j}^{{\rm D}(0)} {^{GG}\!g}_{jk}
-
{^{QQ}\!g}_{jk} {^{QG}\!\gamma}_{k}^{{\rm D}(0)} , \nonumber\\
\label{andimND-GQ}
{^{GQ}\!\gamma}_{jk}^{{\rm ND}(1)}
&=&
\left(
{^{GQ}\!\gamma}_{j}^{{\rm D}(0)} - {^{GQ}\!\gamma}_{k}^{{\rm D}(0)}
\right) d_{jk}
\left(
\beta_0 - {^{QQ}\!\gamma}_{k}^{{\rm D}(0)}
\right)
-
\left(
{^{GG}\!\gamma}_{j}^{{\rm D}(0)} - {^{GG}\!\gamma}_{k}^{{\rm D}(0)}
\right) d_{jk}
{^{GQ}\!\gamma}_{k}^{{\rm D}(0)} \\
&&\qquad\qquad\qquad\qquad\quad +
{^{GQ}\!\gamma}_{j}^{{\rm D}(0)} {^{QQ}\!g}_{jk}
-
{^{GG}\!g}_{jk} {^{GQ}\!\gamma}_{k}^{{\rm D}(0)}
+
\left(
{^{GG}\!\gamma}_{j}^{{\rm D}(0)} - {^{QQ}\!\gamma}_{k}^{{\rm D}(0)}
\right)
{^{GQ}\!g}_{jk} , \nonumber\\
\label{andimND-GG}
{^{GG}\!\gamma}_{jk}^{{\rm ND}(1)}
&=&
\left(
{^{GG}\!\gamma}_{j}^{{\rm D}(0)} - {^{GG}\!\gamma}_{k}^{{\rm D}(0)}
\right)
\left\{
d_{jk}
\left(
\beta_0 - {^{GG}\!\gamma}_{k}^{{\rm D}(0)}
\right)
+ {^{GG}\!g}_{jk}
\right\} \\
&&\qquad\qquad\qquad\qquad\quad -
\left(
{^{GQ}\!\gamma}_{j}^{{\rm D}(0)} - {^{GQ}\!\gamma}_{k}^{{\rm D}(0)}
\right) d_{jk}
{^{QG}\!\gamma}_{k}^{{\rm D}(0)}
-
{^{GQ}\!g}_{jk}{^{QG}\!\gamma}_{k}^{{\rm D}(0)}. \nonumber
\end{eqnarray}
Here the leading order diagonal anomalous dimensions are given by
Eqs.\ (\ref{relations}) and (\ref{LOAD}); the $d$ and $g$ elements
are
\begin{equation}
d_{jk}
= - \frac{1}{2}[ 1 + ( - 1)^{j - k} ]
\frac{(2k + 3)}{(j - k)(j + k + 3)},
\end{equation}
and
\begin{eqnarray}
\label{g-QQ}
{^{QQ}\!g}_{jk}
\!\!\!\!&=&\!\!\!\!
- C_F \left[ 1 + (-1)^{j-k} \right] \theta_{j-2,k}
\frac{( 3 + 2 k )}{(j - k)(j + k + 3)} \\
&\times&\left\{
2 A_{jk} + ( A_{jk} - \psi( j + 2 ) + \psi(1) )
\frac{(j - k)(j + k + 3)}{( k + 1 )( k + 2 )}
\right\} , \\
\label{g-GQ}
{^{GQ}\!g}_{jk}
\!\!\!\!&=&\!\!\!\!
- C_F \left[ 1 + (-1)^{j-k} \right] \theta_{j-2,k}
\frac{1}{6}
\frac{( 3 + 2k )}{( k + 1 ) ( k + 2 )} ,\\
\label{g-GG}
{^{GG}\!g_{jk}}
\!\!\!\!&=&\!\!\!\!
- C_A [ 1 + ( - 1)^{j - k} ] \theta_{j - 2,k}
\frac{( 3 + 2 k )}{(j - k)(j + k + 3)} \\
&\times& \left\{
2 A_{jk} + ( A_{jk} - \psi (j+2) + \psi(1) )
\left[
\frac{\Gamma (j + 4)\Gamma (k)}{\Gamma (j)\Gamma (k + 4)} - 1
\right]
+ 2 (j - k)( j + k + 3 )
\frac{\Gamma (k)}{\Gamma (k + 4)}
\right\} , \nonumber
\end{eqnarray}
respectively. We have introduced here the matrix $\hat A$ whose elements
are defined by
\begin{equation}
A_{jk} = \psi\left( \frac{j + k + 4}{2} \right)
- \psi\left( \frac{j - k}{2} \right)
+ 2 \psi ( j - k ) - \psi ( j + 2 ) - \psi(1) .
\end{equation}

\setcounter{equation}{0}
\renewcommand{\theequation}{\Alph{section}.\arabic{equation}}

%%%%%%%%%%%%%%%%%%%%%%%%%%%%%%%%%%%%%%%%%%%%%%%%%%%%%%%%%%%%%%%%%%%%%
\section{NFPD with support $0 \leq x \leq 1 $.}
\label{CoeffNFPD}
%%%%%%%%%%%%%%%%%%%%%%%%%%%%%%%%%%%%%%%%%%%%%%%%%%%%%%%%%%%%%%%%%%%%%

Here we give a list of formulae for the expansion of non-forward
parton distributions with the support $0 \leq x \leq 1$. The only
difference which arises w.r.t. the results given in the body
of the paper is the argument of the Jacobi polynomials,
$P^{(\alpha, \beta)}_j (2x - 1)$, in the expansion
\begin{equation}
{\mbox{\boldmath$\cal O$}} ( x, \zeta, Q^2 )
= \sum_{j = 0}^{\infty}
\widetilde{\mbox{\boldmath$\cal P$}}_j ( x )
{\mbox{\boldmath$\cal M$}}_j ( \zeta, Q^2 ),
\end{equation}
with elements $\widetilde{\cal P}_j ( x ) =
\frac{w (x | \alpha, \beta )}{n_j (\alpha, \beta)}
P_j^{(\alpha, \beta)} (2x - 1)$ where $n_j (\alpha, \beta)
= \frac{\Gamma (j + \alpha + 1)
\Gamma (j + \beta + 1)}{(2j + \alpha + \beta + 1) j!
\Gamma (j + \alpha + \beta + 1)}$. The Jacobi moments are given by
Eq.\ (\ref{JacobiMom}) provided we will substitute the elements of the
coefficient matrix (\ref{E-matrix}) by (cf. Ref.\ \cite{BelMueNidSch98})
\begin{eqnarray}
E_{jk}^J ( \nu; \alpha, \beta | \zeta)
\!\!\!&=&\!\!\!
(- 1)^{j - k} \theta_{jk}\,
2^{2 \nu - 1} \frac{\Gamma (\nu)}{\Gamma (\frac{1}{2})}
\frac{\Gamma (k + \nu + \frac{1}{2})}{\Gamma (2k + 2\nu)}
\frac{\Gamma (j + \beta + 1)}{\Gamma (k + \beta + 1)}
\frac{\Gamma (j + k + \alpha + \beta + 1)}{\Gamma (j - k + 1)
\Gamma (j + \alpha + \beta + 1 )} \nonumber\\
&&\qquad\qquad\qquad\qquad\times {_3 F_2}
\left.\left(
{ -j + k , j + k + \alpha + \beta + 1 , k + \nu + \frac{1}{2}
\atop
2k + 2\nu + 1 , k + \beta + 1 }
\right| \zeta \right) .
\end{eqnarray}
The results for other classic orthogonal polynomials can be immediately
derived from this equation.

\end{document}